\documentclass[pageno]{jpaper}

\usepackage[normalem]{ulem}
\usepackage{mathptmx} 
\usepackage[sort,nocompress]{cite}
\usepackage[keeplastbox]{flushend}
\usepackage{amsmath,amssymb,amsfonts}
\usepackage{algorithmic}
\usepackage[font={normal,bf}]{caption}
\usepackage{textcomp}
\usepackage{fancyhdr}
\usepackage{microtype}
\usepackage{color, soul}
\usepackage{xspace}
\usepackage{multirow}
\usepackage{graphicx}
\usepackage{booktabs}
\usepackage{pifont}
\usepackage{float}
\usepackage{xcolor}
\usepackage{makecell}
\usepackage{tikz}
\usepackage{xcolor}
\usepackage{enumitem}
\usepackage{listings}
\usepackage{color}
\usepackage[symbol]{footmisc}
\usepackage{url}
\usepackage{xurl}
\usepackage{authblk} 
\newcommand{\ignore}[1]{}

\newcommand{\ju}[1]{\textcolor{purple}{#1}}

\definecolor{darkgreen}{rgb}{0.0, 0.7, 0.13}
\newcommand{\cmark}{\color{darkgreen}\ding{51}}%
\newcommand{\xmark}{\color{red}\ding{55}}%
\newcommand{\sys}[0]{\textsc{Stockade}\xspace}
\newcommand{\Sys}[0]{\textsc{Stockade}\xspace}
\newcommand{\se}[0]{bi-enclave\xspace}
\newcommand{\ses}[0]{bi-enclaves\xspace}

\newcommand{\Se}[0]{Bi-enclave\xspace}
\soulregister\cite7
\soulregister\ref7
\soulregister\pageref7
\soulregister\textit7
\soulregister\textbf7

\definecolor{dkgreen}{rgb}{0,0.6,0}
\definecolor{dkblue}{rgb}{0,0,0.6}
\definecolor{grgray}{rgb}{0.55,0.55,0.55}
\definecolor{gray}{rgb}{0.5,0.5,0.5}
\definecolor{mauve}{rgb}{0.58,0,0.82}

\begin{document}
\title{Stockade: Hardware Hardening for Distributed Trusted Sandboxes}

\author[1]{Joongun Park}
\author[1]{Seughyo Kang}
\author[1]{Sanghyun Lee}
\author[2]{Taehoon Kim}
\author[1]{\\Jongse Park}
\author[1]{Yongjin Kwon}
\author[1]{Jaehyuk Huh}
\affil[1]{School of Computing, KAIST}
\affil[2]{ETRI}

\date{}
\maketitle
\thispagestyle{empty}

\begin{abstract}

Recent studies showed that a cloud application consists of multiple distributed modules provided
by mutually distrustful parties.
For trusted services, such applications can use trusted execution environments (TEEs) communicating 
through software-encrypted memory channels.
Such an emerging TEE execution model requires a new type of bi-directional protection: 
protecting the rest of the system from the enclave module with sandboxing
and protecting the enclave module from third-party modules and the operating system.
However, the current TEE model cannot efficiently represent such distributed sandbox applications.
To overcome the lack of hardware supports, this paper proposes an extended TEE model
called {\it \sys}, which supports distributed sandboxes hardened by hardware.
{\it \Sys} proposes new three key techniques. First, it extends the hardware-based
memory isolation in SGX to confine a user software module only within its TEE (enclave). Second,
it proposes a trusted monitor enclave that filters and validates systems calls from enclaves.
Finally, it allows hardware-protected memory sharing between a pair of enclaves for
efficient protected communication without software-based encryption.
Using an emulated SGX platform with the proposed extensions, this paper shows that
distributed sandbox applications can be effectively supported with small changes
of SGX hardware.

\end{abstract}


\section{Introduction}

Hardware-based trusted execution environments (TEEs) enabled the strong isolation of execution contexts
in remote clouds, even when the servers are exposed to potential vulnerability in privileged software and
physical attacks. Among recent TEE supports, Intel Software Guard Extension (SGX), a commercial incarnation of TEEs,
provides isolated execution environments called {\it enclaves} protected by the CPU hardware.
The CPU hardware isolates each enclave from the operating system. Its code and data are
encrypted and integrity-verified while they reside in the external DRAM.

\begin{figure}[t]
\centering
\includegraphics[width=8.4cm]{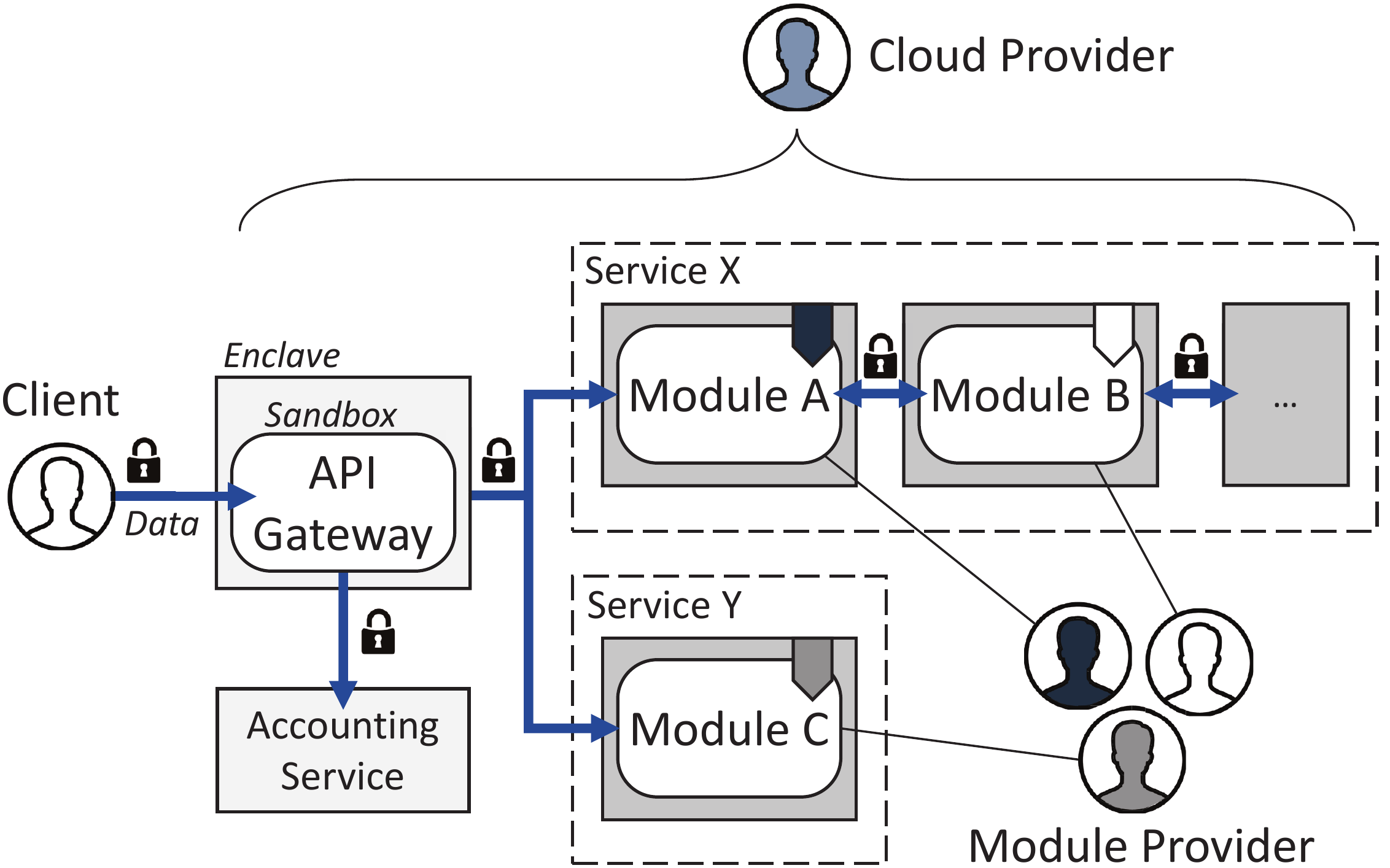}
\caption{A distributed sandboxes with SGX. Gray boxes are enclaves running modules from different providers}
\label{fig:application}
\end{figure}

The introduction of commercially available TEEs has been accelerating the exploration of application scenarios
utilizing their strong isolation capability.
One important cloud-oriented scenario is to provide function-as-a-service or software-as-a-service on clouds, running a function or software
in each enclave~\cite{openlambda,amazon_lambda, google_app_engine}. In such applications, it is critical not only to protect user-provided functions
from the potentially vulnerable
cloud system but also to secure the hosting cloud system by sandboxing the user-provided functions or software, as they
cannot be fully trusted from the perspective of the hosting system.
Besides, an application task is composed of multiple functions communicating with each other~\cite{ryoan}.
Figure~\ref{fig:application} presents such a distributed sandboxed application.
Such an application consists of software modules from multiple software providers, which may not entirely trust
the other providers. With multiple participants, each module must be protected from other modules or the hosting system,
and modules must also be confined to prevent any exploitation of system vulnerability.
A key software technique for such distributed secure applications is {\it software sandboxing} which prevents
the codes in an enclave from accessing the memory beyond the protected enclave memory and validates system calls.

The distributed sandboxed applications reveal the limitations of the current SGX model.
First, the codes inside an enclave can freely access the remaining untrusted memory of the process. Such uni-directional
protection can endanger the rest of the system if the enclave code is malicious. To address
such vulnerability, the prior study proposed to employ a heavy software sandboxing running with user codes inside
an enclave~\cite{ryoan,occlum,enclavedom,Chancel,AccTEE}.
Second, enclaves require to use operating system services via system calls, but the secure interaction via system calls must
be considered. Not only such system call requests must be verified to protect the hosting system~\cite{SGXJail},
but return values must be checked to prevent Iago attacks against the enclave~\cite{emilia,PANOPLY,graphene-sgx,scone}. 
Finally, the communication channel among enclaves is not provided by the hardware mechanism.
For secure inter-enclave communication, a pair of enclaves must share an untrusted memory region,
and each message must be encrypted and integrity-protected by the software running inside the enclaves.
Such software-based encrypted communication not only increases the communication latency but also
can cause a vulnerability~\cite{TOCTOU, PANOPLY}.

To overcome the limitations of the current TEE model, this study proposes an extension of the enclave model,
called {\it \sys}. \sys provides efficient hardware-supported solutions for the three
limitations. First, instead of using software-based sandboxing, \sys blocks accesses from enclaves
to the untrusted world. We call the sandboxed enclave \textit{\se}.
By simply extending the pre-existing memory validation mechanism in SGX hardware,
a \se can not only be protected from the untrusted world but also be prevented from accessing the untrusted
context. Such bi-directional isolation enables solid sandboxing support for each bi-enclave without any extra
software layer.

The second mechanism is to provide a hardened interaction between a \se and the operating system.
The interaction of the \se and operating system can be forced to go through the monitor enclave
to process the system calls only if they are valid.
The key difference from the prior approaches~\cite{ryoan,occlum,enclavedom,SGXJail,Chancel, AccTEE} is that the monitor is
isolated both from the \se and from the operating system, which provides stronger protection for 
the system call verification and return value validation.
The codes running in the monitor enclave are attested by both the \se and operating system,
providing verified monitoring operations by the two entities.
With the neutral monitor enclave, \sys can provide a temper-proof accounting service
of system resources such as file I/Os and network usages, as both the cloud users and providers can
trust the monitor enclave. 

The final mechanism allows sharing of trusted memory pages between two enclaves. The hardware provides
an interface for sharing the protected pages between two enclaves, and the memory isolation mechanism is extended
to allow two enclaves to access the shared pages.
By sharing the hardware-protected pages, the communication between the enclaves does not require costly software-based
encryption and integrity protection.

To show the effectiveness of the new enclave extensions, we ported several application scenarios
on an emulated SGX runtime with the extended interface. The experimental results show that
minor hardware extensions can improve the efficiency and security of distributed sandbox applications
on clouds.
Compared to the prior SW-based sandboxing, it provides 1.4$\sim$19.5\% performance improvements.
This study hardens the distributed sandbox applications with hardware extensions. To the best of
our knowledge, it is the first study to extend the execution model and hardware for bi-directional protection
with the protected system call monitor. The new contributions of this paper are as follows:

\begin{itemize}

\item It proposes bi-directional isolation between an enclave and its untrusted environment.
      The design shows that a simple extension of the existing memory access control mechanism in
      SGX can provide efficient isolation for both ways.

\item It proposes a hardware-protected monitoring mechanism for handling system call filtering and
      accounting operations for each enclave.

\item It proposes a shared trusted memory between two enclaves. With a careful design, a designated
      part of the protected memory of an enclave can be shared with the other enclave.

\end{itemize}

The rest of the paper is organized as follows.
Section~\ref{sec:background} presents the background of distributed sandbox applications.
Section~\ref{sec:motivation} discusses the motivations of three extensions, and
Section~\ref{sec:related} discusses the related works. 
Section~\ref{sec:design} presents the proposed hardware extensions.
Section~\ref{sec:discussion} presents the security analysis, and
Section~\ref{sec:evaluation} provides four application scenarios using \se and their performance on an emulated
SGX runtime.
Section~\ref{sec:conclusion} concludes the paper.



\section{Background}
\label{sec:background}

\subsection{Intel Software Guard Extensions (SGX)}

Intel SGX provides a user-level trusted execution environment called an {\it enclave}.
The context of an enclave is protected by the hardware mechanism. 
The protected memory region of enclaves is created in Enclave Page Cache (EPC).
Part of physical memory, Processor Reserved Memory (PRM), is reserved for
SGX and is protected by the hardware memory encryption engine~(MEE).
PRM contains the EPC pages in addition to other security meta-data for SGX.
Although EPC pages are in the external DRAM, their confidentiality and integrity
are guaranteed under direct physical attacks on DRAM and system interconnection components.
The attestation support allows a user to verify the identity and measured digest of an enclave and platform setting
where the enclave runs.


The memory isolation for each enclave is done during the address translation step for each
memory access. A mode transition between the enclave mode and untrusted mode requires
flushing Translation Lookaside Buffers (TLBs). For each TLB miss, the validity of
access is verified by the CPU hardware logic. A key internal data structure
for verification is Enclave Page Cache Map (EPCM) which is stored in PRM.
An EPCM entry has information about a physical page that belongs to the EPC region.
It contains the owner's enclave ID and its virtual address in the enclave memory space, in addition to other status information.
Even though page tables are still managed and updated by the operating system, the EPCM table is accessible
only by the hardware, and the page table entry for EPC can be verified using EPCM.
The crucial invariant for the correctness of memory isolation is that
{\it TLB must contain only verified translations}.

SGX controls enclave through a set of instructions. After an enclave is created,
\textit{EINIT} initializes it to be ready for protected execution.
The virtual address of protected memory region for an enclave is fixed during
the initialization of the enclave. 
The virtual address range for an enclave should be a single contiguous region called Enclave Linear Address Range (ELRANGE).
The context information of an enclave is stored in its SGX Enclave Control Structures (SECS).
SECS are allocated in EPC pages for its safety against the malicious operating system.
SGX includes instructions for switching modes between
enclave context and unprotected context: \textit{EENTER} to enter enclave mode, and \textit{EEXIT} to exit enclave mode.




\subsection{Sandboxing}
Sandboxing confines an application in its own environment.
By isolating untrusted applications, sandboxing protects the kernel and host environment against potential attacks
from the applications.
Sandboxing is widely adopted for runtime protection against third-party applications,
such as web browsers running plugins written by unauthorized developers~\cite{chrome_sandbox,mozilla_sandbox},
and testbeds for third-party developers migrating their applications to the production system~\cite{google_app_engine,microsoft_sandbox,amazon_sandbox,yahoo_sandbox,paypal_sandbox_1,paypal_sandbox_2,ebay_sandbox}.

An application running in a sandbox must not be allowed to directly access the memory outside of the sandbox.
In addition, the application control should never reach beyond the designated sandbox, neither directly nor indirectly during its runtime.
To provide sandboxing, fault isolation confines control transfer and data access within a sandbox,
and system call filtering validates system call requests from sandbox applications.

\noindent
{\bf Fault Isolation:}
Fault isolation provides a logically isolated compartment by enforcing its confinement
policy on memory and control transfer. Software-based fault isolation provides such confinement by
binary instrumentation or compiler support~\cite{SFI,fastSFI,NaCl,adaptingSFI,webAssembly}.
Using binary instrumentation, Google Native Client (NaCl) restricts memory accesses from untrusted applications,
by masking target addresses with memory boundary before the binary execution.
Such software-based isolation needs to execute extra instructions for \ju{access} validation, adding performance overheads.
In addition, the instruction-based bound checking is potentially vulnerable to the Spectre attacks~\cite{swivel,spectre,meltdown,systematic,spectrereturns}.
Other fault isolation techniques rely on CPU hardware supports for confinement~\cite{ARMlock,flexdroid,SGXJail,enclavedom}.
With hardware supports such as Intel Memory Protection Keys (MPK)~\cite{libmpk}, or ARM Domain~\cite{armdomain},
they provide sandboxes to separate modules from each other. However, the current MPK uses page tables 
to track memory domains, and thus the domain information can be changed by OS.



\noindent
{\bf System Call Monitoring:}
In addition to the memory access control, the interaction with the operating system must also be regulated by sandboxing.
Although the operating system is protected with privilege separation and system call interfaces, system vulnerabilities via system calls
have been continuously reported~\cite{CVE-2019-2054,CVE-2020-17087,shellshock,CVE-2019-3969}.
A naive way to alleviate this problem is not allowing untrusted applications to make any system calls.
However, many real-world applications are relying on system call interfaces such as POSIX to use network supports and file management.
Therefore, the sandbox must provide controlled system functionalities by verifying system calls from the untrusted application.
Seccomp-bpf~\cite{seccomp} interposes system call requests by filtering system call with ID and arguments.
In addition to filtering system calls, by manipulating return values of system calls, 
a malicious operating system can leak the application's secret or break the execution integrity
known as the Iago Attack~\cite{Iago}. To prevent Iago attacks, return values also need to be validated~\cite{Sego,Inktag}.

\subsection{Cloud Applications and Trusted Execution}

Cloud services have evolved to use a more complex task model, where many different software modules are interacting with
each other. A single cloud application may rely on multiple modules from different parties.
As shown in Figure~\ref{fig:application}, a recent advancement of function-as-a-service or software-as-a-service has enabled
cloud applications to be composed of small functions or modules.
Each module is implemented by a different party, and thus, its trustworthiness
is not fully guaranteed from the perspective of the other module providers.
In addition, the cloud provider must protect the system from modules and clients.
To support such new application scenarios, the trusted execution model needs to evolve.

Recent studies investigated applying trusted computing to such distributed cloud applications consisting of multiple modules with
different providers~\cite{ryoan,sfaas,clemmys,multi-domains,awslambda}. 
They proposed to run each module in an enclave, but additionally sandboxing
is combined with the module running in an enclave.
The sandbox library blocks access to untrusted memory from an enclave, which confines the access boundary of an enclave only to
its own EPC region.
With such software sandboxing, it attempts to prevent the functions in an enclave from exploiting potential vulnerabilities of the system.
In addition, it uses software-encrypted communication via shared memory between enclaves to allow the coordination and data transfer
of multiple modules.

Such cloud applications require extensions of the current SGX model:
First, TEE not only needs to protect the context in an enclave but also must confine accesses from enclaves, if necessary.
Second, a software module in an enclave often needs to access the system resource via system calls. How to control the system call
access must also be considered in the SGX model.
Third, multiple software modules must efficiently interact with each other.
However, the current SGX is not designed for facilitating inter-enclave communication.


\section{Motivation}
\label{sec:motivation}

\subsection{Bi-directional Isolation with Enclave}

In distributed sandboxed applications, an enclave execution must be protected,
but it must also be prevented from accessing memory beyond its own EPC region.
In the current SGX model, the in-enclave execution is freely allowed to access the rest of
its process memory, which is confined only by the operating system.
There are two different ways of providing confinement supports for the current SGX enclave model.

\begin{figure}[t]
\centering
\includegraphics[width=8.5cm]{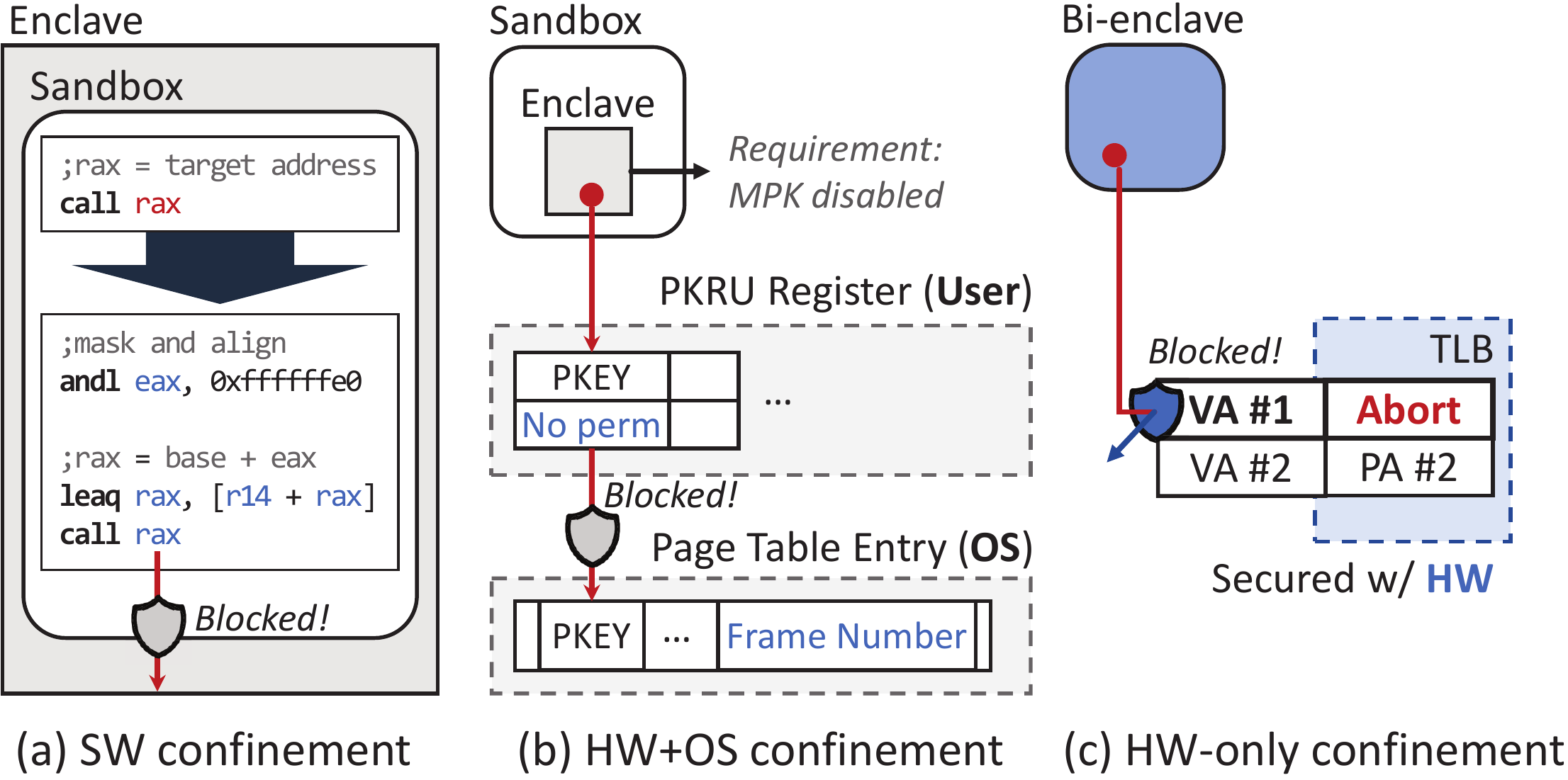}
    \caption{Three confinement approaches: SW, HW+OS (MPK), and HW-only (Bi-enclave)}
\label{fig:confinement}
\end{figure}

The software-based confinement for the enclave is to include the instrumented application binary and sandbox library codes
together in each enclave. In such an approach, the sandbox library, as well as application binary, must be trusted.
Figure~\ref{fig:confinement} (a) describes the software-based approach (SW).
Combining the sandbox runtime with the application binary in a single enclave increases
the Trusted Computing Base (TCB) of the enclave. When a vulnerability exists in the sandbox library~\cite{CVE-2019-2054},
the application code can potentially exploit the vulnerability to bypass the confinement.
In addition to the increased TCB, the memory access occurring in the enclave must be verified by instrumented instructions, causing
extra performance costs.
A recent study~\cite{Chancel} reports that the software-based confinement incurs a slowdown of an average of 12.43\%,
up to 24.89\% compared to native execution because it requires 23.52\% more instructions. 

Recent hardware supports for memory confinement such as Intel Memory Protection Keys (MPK) can mitigate
the weaknesses of software-only approaches. However, the current hardware-assisted mechanism relies
on page tables for tracking the isolated memory domains. Since page tables can be modified by
any privileged access, the confinement assumes that the operating system is trusted.
Figure~\ref{fig:confinement} (b) shows the hardware-assisted approach with MPK (HW+OS).
A limitation of MPK is that 
the PKRU registers which defines the permission for each domain are user-accessible.
Therefore, the user application in an enclave must be verified not to update the PKRU registers.
In addition, the current HW+OS approach relies on the security of page tables.
However, recent studies showed that page tables can be vulnerable to various attacks 
including rowhammer attacks~\cite{Cross_VM_Row_Hammer_Attack, Seaborn, Cheng2018StillHA, Drammer, Rowhammer.js}.


To overcome the limitations of the current software-only and hardware-assisted sandbox designs, this paper
proposes to extend the SGX memory access control mechanism.
Unlike MPK, it does not use page tables to store critical domain information.
Figure~\ref{fig:confinement} (c) shows the pure hardware-enforced approach of \Sys (HW-only).


\subsection{OS Interactions with Bi-enclave}

\begin{figure}[t]
\centering
\includegraphics[width=8.5cm]{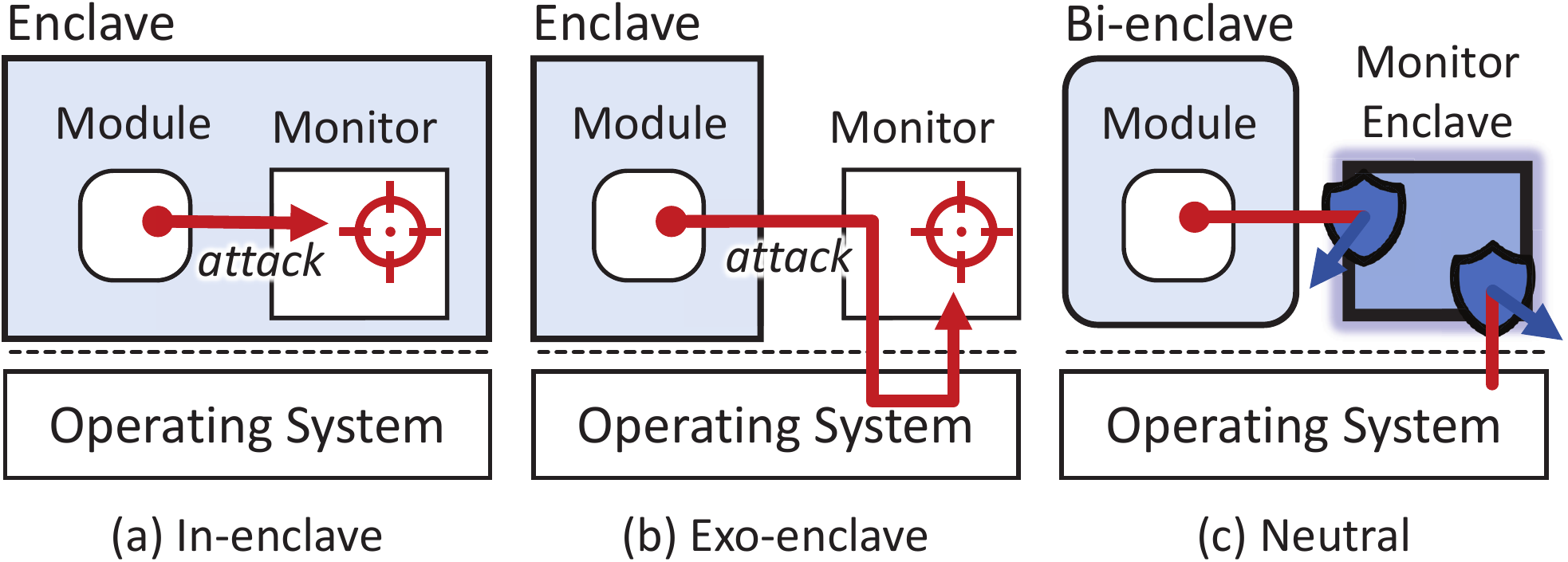}
\caption{Three syscall monitor approaches: in-enclave, exo-enclave, and neutral enclave}
\label{fig:sandbox_comp}
\end{figure}

A \se is prohibited to access the memory outside of its own EPC, but it should be allowed
to issue system call requests if the system call requests can be verified for their safety.
Therefore, to provide fully functioning sandbox enclaves, it is necessary to support a safe
mechanism to verify system call requests and forward the filtered requests to the operating system.
In addition, the returned value from the untrusted operating system may need to be checked.
On the contrary, in the prior distributed sandbox approach~\cite{ryoan}, a confined module is not allowed
to use any system call directly.

There are two different approaches to provide system call services to enclave execution:
system call emulation and system call delegation.
First, the system call emulation approach imports the entire library OS~\cite{drawbridge,graphene,lkl} and C standard libraries~\cite{musl}
inside an enclave~\cite{Haven, graphene-sgx, scone, occlum, enclavedom, sgx-lkl}.
With the intra-enclave libOS, porting efforts for existing applications to SGX is minimized.
In addition, with a carefully designed shim layer,
it helps to minimize the exposed interfaces between the host and enclave
which is the main attack surface of the enclave~\cite{CoinAttack}.
However, this approach adds the entire software stack within an enclave, increasing the TCB of an enclave significantly.
Figure~\ref{fig:sandbox_comp} (a) shows the in-enclave monitor approach. It assumes that the monitor code can
be completely isolated from the application module by a software confinement layer. However, as discussed in the prior subsection,
the vulnerability in SW-only confinement may not guarantee the protection of the monitor.

Second, the system call delegation approach relies on the underlying OS itself, thereby reducing TCB drastically~\cite{SGXJail,PANOPLY}.
Rather than including a large libOS stack within each enclave, it includes a much smaller codebase for system call interposition and verification after execution.
It delegates system calls to the non-enclave mode and performs its system call.
The returned results from the system call are verified inside the enclave to prevent malicious intervention.
In both cases, the enclave must be able to interact with the untrusted context for system calls.
Figure~\ref{fig:sandbox_comp} (b) shows the exo-enclave monitor approach.
It assumes that the monitor code exists as a process in unprotected environments accessible from the OS kernel.
Therefore, this approach is vulnerable to attacks like privilege escalation~\cite{privilege_escalation_1, shellshock, Process-injection}.

%

\renewcommand{\arraystretch}{0.8}
\begin{table*}[t]
\centering
\resizebox{0.99\textwidth}{!}{%
\small
\begin{tabular}{l|cc|cc|cccc|cc}
\toprule
    & \multicolumn{2}{c|}{\textbf{Confinement}} & \multicolumn{2}{c|}{\textbf{System call filter}}\hspace{0.5cm} & \multicolumn{4}{c|}{\textbf{Protection against}} & \multicolumn{2}{c}{\textbf{Multi module support}} \\  
\cmidrule{2-11}
    \textbf{Related Work} & \multicolumn{1}{c}{Type} & \multicolumn{1}{c|}{Method} & \multicolumn{1}{c}{Type} & \multicolumn{1}{c|}{Method} \hspace{0.5cm}
	& \begin{tabular}[l]{@{}l@{}}SW Sandbox\\Vulnerability\end{tabular} & \begin{tabular}[l]{@{}c@{}}Privilege\\Escalation\end{tabular} 
	& \begin{tabular}[c]{@{}c@{}}PTE\\Corruption\end{tabular} & Iago attack 
	& \begin{tabular}[l]{@{}l@{}}3rd Party\\Module\end{tabular} & \begin{tabular}[l]{@{}l@{}}Protected\\Channel\end{tabular}\\
\midrule
Ryoan~\cite{ryoan} 	        & SW 	& Native Client 		& In-enclave  & \multicolumn{1}{c|}{Libc} \hspace{0.5cm} & \hspace{-0.5cm} \xmark & \cmark & \cmark & \cmark & \cmark & Inter-enclave (SW)\\
Chancel~\cite{Chancel}      & SW 	& LLVM 	& In-enclave  & \multicolumn{1}{c|}{Libc} \hspace{0.5cm} & \hspace{-0.5cm} \xmark & \cmark & \cmark & \cmark & 					\xmark & Intra-enclave\\
AccTEE~\cite{AccTEE} 	    & SW 	& WebAssembly 	& In-enclave  & \multicolumn{1}{c|}{LibOS} \hspace{0.5cm} & \hspace{-0.5cm} \xmark & \cmark & \cmark & \xmark & \xmark & \xmark\\
Occlum~\cite{occlum} 	    & HW+OS & MPX~\footnote[3]{}	 	& In-enclave  & \multicolumn{1}{c|}{LibOS} \hspace{0.5cm} & \hspace{-0.5cm} \xmark & \cmark & \cmark & \xmark & \xmark & Intra-enclave\\
EnclaveDom~\cite{enclavedom}& HW+OS & MPK~\footnote[2]{}	& In-enclave  & \multicolumn{1}{c|}{LibOS} \hspace{0.5cm} & \hspace{-0.5cm} \xmark & \cmark & \xmark & \xmark & \cmark & Intra-enclave\\
SGXJail~\cite{SGXJail} 	    & HW+OS & MPK~\footnote[2]{}	& Exo-enclave & \multicolumn{1}{c|}{Seccomp} \hspace{0.5cm} & \hspace{-0.5cm} \cmark & \xmark & \xmark & \xmark & \xmark & \xmark\\
\midrule
\Sys                        & \textbf{HW-only} & \textbf{SGX} & \textbf{Neutral} & \multicolumn{1}{c|}{\textbf{Monitor}} \hspace{0.5cm} & \hspace{-0.5cm} \cmark & \cmark & \cmark & \cmark & \cmark& \textbf{Inter-enclave (HW)}\\
\bottomrule
\multicolumn{11}{l}{} \\
\multicolumn{1}{l}{\cmark : Considered / Secure} & \multicolumn{2}{l}{\xmark : Not considered / Vulnerable} & \multicolumn{1}{l}{\footnote[3]{}Deprecated~\cite{mpx,intelmpx}} &\multicolumn{1}{l}{\footnote[2]{}HW modification required}
\end{tabular}
}
\caption{Comparing \sys to prior work}
\label{tab:relatedwork}
\end{table*}
\renewcommand{\arraystretch}{1.0}

To allow the controlled interaction from the sandbox enclave and operating system,
we propose to add a {\it monitor enclave} that can be coupled with one or more \ses.
A monitor enclave is a conventional enclave, and thus it can jump to the system call function in
the untrusted region. Between the \se and monitor enclave, a protected memory channel
is created, and the \se sends a request to the monitor enclave.
As shown in figure~\ref{fig:sandbox_comp} (c), \sys takes a different approach to others.
\Sys locates the monitor in a position-neutral enclave.
In the approach, the monitor is protected both from the user enclave and OS kernel.

Sandboxes using system call delegation have to consider races between sandboxes~\cite{ostia}.
In addition, a prior study~\cite{emilia} observed that a Iago attack can occur across multiple components,
and thus checking the return value within each enclave individually is not enough to prevent such an attack.
In \Sys, the monitor can track global states among bi-enclaves thus can prevent Iago attacks against connected bi-enclaves.
In addition, the design helps developer not to add Iago attack protection in every bi-enclave.

\noindent
{\bf Tamper-proof resource accounting: }
One of the requirements for the trusted cloud service is tamper-proof resource accounting~\cite{HRA, T-lease, AccTEE}.
For each user, the system resource usages must be securely tracked and reported.
To support such tamper-proof accounting which can be trusted by both cloud users and service providers,
it is necessary to track system resource usages by a mutually trusted entity. 
As a monitor enclave can be isolated from both users and OS, it can act as a neutral accountant, recording
the file and network I/Os.

\subsection{Interactions between Enclaves}

\begin{figure}[t]
\centering
\includegraphics[width=8cm]{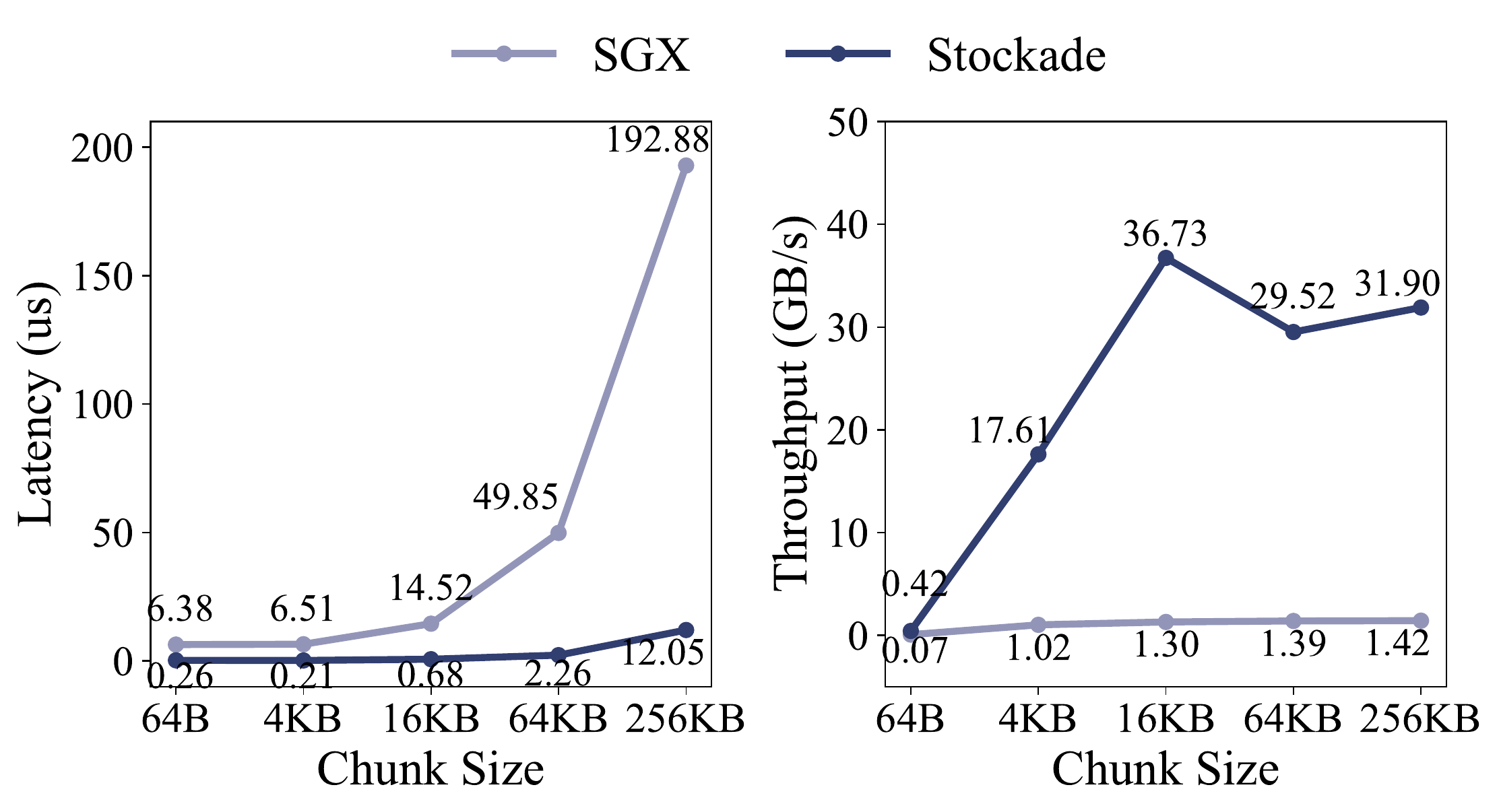}
\caption{Comparing inter-enclave communication between SGX and \Sys}
\label{fig:perf_comm}
\vspace{-0.05in}
\end{figure}

Intel SGX supports function-like interfaces, ecall and ocall, between an enclave and unprotected side for context switching.
However, SGX does not provide APIs for inter-enclave communication.
The prior work used a software-based reliable messaging mechanism with shared untrusted memory between enclaves~\cite{ryoan}.
The messages are encrypted with a shared key. Message authentication code and counters ensure the integrity and freshness of the message.
The protection mechanism can be further improved by padding or truncating the messages for making the attacker hard to guess the original message size.
The software module for cryptography must be trusted and verified, as it is included in the enclave binary.
However, Panoply~\cite{PANOPLY} showed that an attacker can abort the application silently and make the application misjudge its state
by dropping messages even the communication channel is encrypted.

In this study, we use a similar message-based interaction model, but propose to extend the hardware memory access control to allow
the protected memory channel.
We propose a new secure memory sharing model between two enclaves, which uses a designated part of EPC memory region.
The memory region is accessible only by the two enclaves, and the hardware memory access control and memory encryption engine protect it.
It eliminates the need for software-based memory encryption, which increases the latency and may cause
potential vulnerabilities as exploited by Panoply. 
Figure~\ref{fig:perf_comm} shows the communication latency and bandwidth between the current SGX and proposed \se communication.
The SGX model uses software-based encryption(AES-GCM) via untrusted shared memory, while the \se configuration directly shares the hardware
protected memory. As shown in the figure, the latency for transferring data is much higher with SGX than \sys, and the difference
increases as the chunk size increases. In addition, the throughput of \sys communication exhibits much higher bandwidth
than the software-encrypted channel.

\subsection{Comparison to the Prior Work}
\label{sec:related} 

There have been several prior work for sandboxing within a TEE.
Minibox~\cite{Minibox} presents the first two-way sandbox for native x86 code, providing secure file I/O and Iago attack protection.
Ryoan~\cite{ryoan} modifies Native Client~\cite{NaCl} for its software sandboxing.
Similar to \Sys, Ryoan decomposed a cloud application into distributed enclaves with the software sandboxing and SW-encrypted channels.
Chancel~\cite{Chancel} proposes multi-client software fault isolation through binary instrumentation and read-only shared memory between threads.
It supports multiple isolated threads within an enclave.
Several studies accommodate hardware features (Intel MPX or MPK) under software control for multi-domain SFI scheme
~\cite{occlum, mdSFI, enclavedom, Chancel}.
Occlum~\cite{occlum} and Enclavedom~\cite{enclavedom} provide isolated compartments within a sandboxed enclave using Intel MPX or MPK.
SGXJail~\cite{SGXJail} isolates an enclave instance for each process, 
and the system call filtering is provided by the seccomp filters in each process.


Table~\ref{tab:relatedwork} presents the confinement mechanism and monitor locations of the prior works compared to \Sys.
Only \Sys can provide comprehensive protection against four attack types, SW sandbox vulnerability, privilege escalation, rowhammer on PTE, and Iago attack.
In addition, \Sys allows secure hardware-protected inter-enclave communication. 

\noindent
{\bf Other related work: }
There are studies~\cite{Haven,scone,graphene-sgx,occlum} provide trusted library OSes running in the enclave, and enable unmodified application execution in the enclave. 
Panoply~\cite{PANOPLY} reduces TCB by delegates syscalls to OS and verify later.
Nested Enclave~\cite{NestedEnclave} presents static sharing enclave and communication via the outer enclave. 
\Sys provides dynamic EPC sharing with page granularity.

\subsection{Threat model}
\label{subsec:threatmodel}

\Sys shares the basic threat model and trusted computing base (TCB) of SGX.
The SGX-enabled processor package is trusted.
Privileged software such as the operating system and hypervisor can be
compromised by its vulnerability or any person who obtains the privilege permission.
Moreover, attackers can wield direct physical attacks on on-board interconnections
and external DRAM. 


A different assumption of this work over SGX is in the trustworthiness of software modules running
in enclaves. In SGX, a software module running in an enclave is trusted, and potential
attacks from the module itself to the rest of the system are not considered.
We assume that each module does not fully trust the other modules, even when
they are used together to build an application. 
In our model, the code running in the monitor enclave is trusted, and
the monitor enclave and a \se are mutually protected from each other.


\noindent
{\bf Out of scope:}
Architecture defects~\cite{Foreshadow,foreshadow-ng,spectre,meltdown,tlbleed}, side channel attacks~\cite{telling,inferring,sgxcacheattack,controlled,guard,survey,microscope}, and availability are not considered in this work.
For such attacks, prior patches~\cite{foreshadow_patch} and protections~\cite{tsgx,preventing,raccoon,thwarting,obfuscuro,sgx-shield,CoinAttack} 
can be used as orthogonal measures.
\sys does not support resiliency to code reuse attacks\mbox{\cite{guard,hacking}} and arbitrary API invocation (e.g. COIN attack\cite{CoinAttack}).

\section{Architecture}
\label{sec:design}

\subsection{Overview}
\label{subsec:overview}

\begin{figure}[t]
\centering
\includegraphics[width=8.5cm]{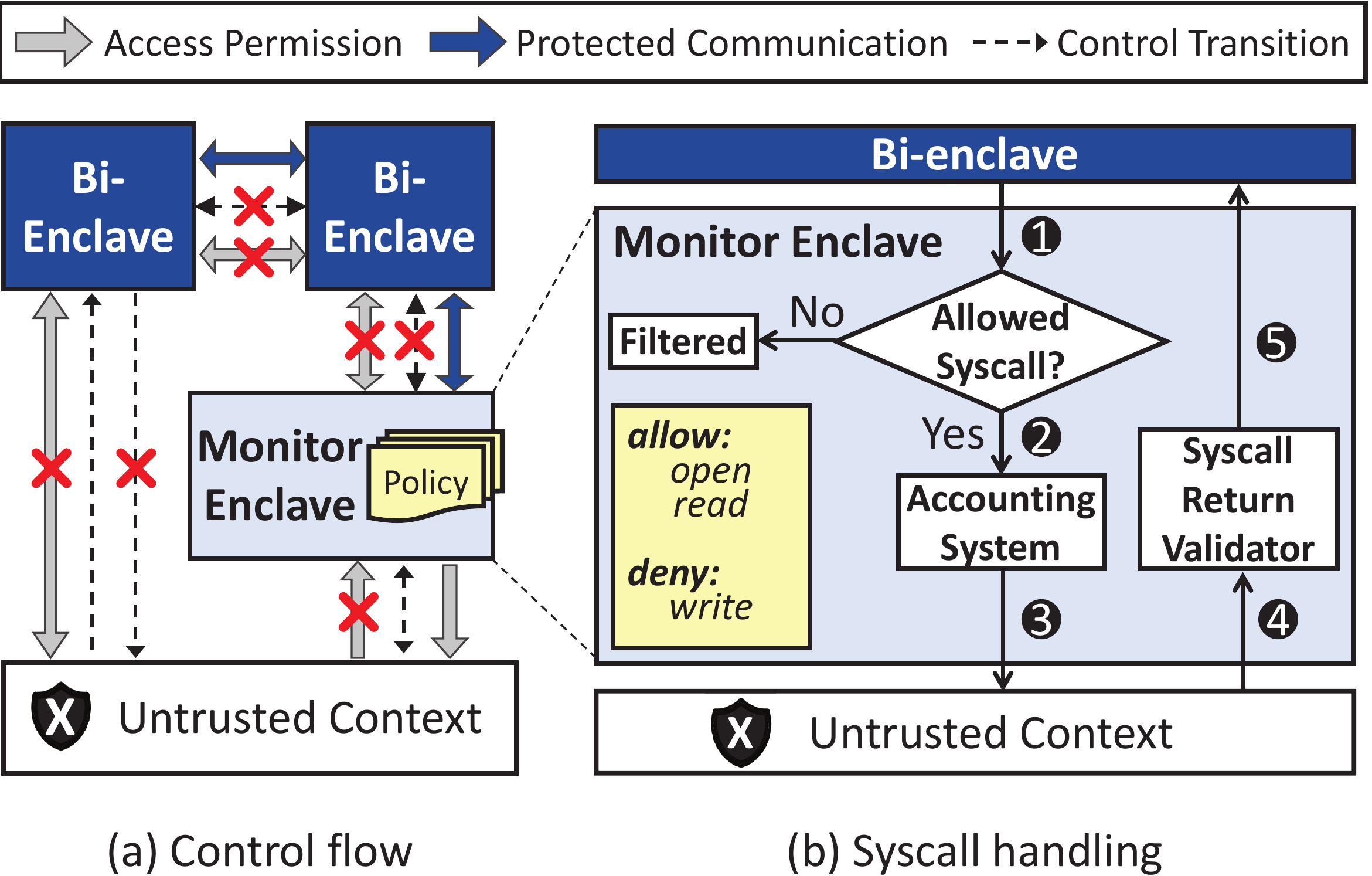}
\vspace{-0.1in}
\caption{\Sys architecture. The control flow from the untrusted context to \se is only allowed at launching}
\label{fig:secure_sandbox}
\vspace{-0.1in}
\end{figure}

Figure~\ref{fig:secure_sandbox} presents \sys's distributed sandboxing model.
\Sys architecture has two types of enclave, {\it \se} and {\it monitor enclave} as shown in (a).
\Se inherits all the properties of SGX enclave, and additionally,
a \se blocks all accesses to the non-enclave memory with the hardware support.
The code running in a \se is not allowed to read, write, and execute
contents outside the \se memory.
In addition, the control of a \se cannot be directly transferred to the non-enclave context, but
it must go through the monitor enclave to interact with the rest of the system.
\Sys allows the monitor enclave to communicate with the operating system (OS).

The monitor enclave works as a proxy to communicate the operating system.
To interact with the operating system, a \se has to establish a secure shared memory channel to a monitor enclave.
For communication with other enclaves, the same secure shared memory channel is supported.
As shown in Figure~\ref{fig:secure_sandbox} (b), with the monitor enclave attached to a \se, it delegates system call to OS.
A monitor enclave verifies system calls based on a given profile and validates the return values of system calls
to prevent known Iago attacks.
In addition, the monitor enclave can track system call usage records in a mutually trusted way.

\noindent
{\bf Application model:} 
In \sys model, a service consists of one or more mutually distrustful modules.
Each module is enclosed in a \se, and mutually distrustful modules do not reside in the same \se.
With the protection boundary, both \se and monitor can have multiple threads.

\renewcommand{\arraystretch}{0.8}
\begin{table}[t]
\centering
\resizebox{8.8cm}{!}{%
\small
\begin{tabular}{l|l}
\toprule
\multicolumn{1}{l|}{\textbf{Target}} & \textbf{Description} \\
\midrule
SECS* 						& New field (1 bit) for \se flag ($\S$\ref{subsec:overview}) \\
EPCM entry*					& New field (52 bits) for phsycal address of Co-owner's SECS ($\S$\ref{subsec: memory_protection}) \\
\midrule
EINIT\footnote[2]{}			& Set \se flag in SECS on initialization ($\S$\ref{subsec:overview})\\ 
EEXIT\footnote[2]{}	 		& Abort EEXIT when \se flag is set in SECS ($\S$\ref{subsec: memory_protection})\\
ESADD\footnote[2]{}			& Establish shared EPC to target enclave (Ring 3) ($\S$\ref{subsec:shared_epc})\\
ESACCEPT\footnote[2]{}	 	& Accept shareable EPC from ESADD (Ring 3) ($\S$\ref{subsec:shared_epc})\\
\midrule
TLB\footnote[3]{}	 			& Fill entry on Co-onwer access / Abort on forbidden access ($\S$\ref{subsec: memory_protection})\\
\bottomrule
\multicolumn{2}{l}{} \\
\multicolumn{2}{l}{*Data Structure \quad \footnote[2]{}Instruction \quad \footnote[3]{}Access checks on TLB miss} \\
\end{tabular}
}
\vspace{-0.1in}
\caption{Summary of hardware changes}
\label{tab:hardware_changes}
\vspace{-0.1in}
\end{table}
\renewcommand{\arraystretch}{1.0}

\noindent
\textbf{Hardware changes:}
The SGX security features are mostly implemented in microcode which incurs much less implementation overheads than CPU circuitry~\cite{sgx-explained}
Therefore, the majority of modifications for \sys on data structures, instructions, and access control, can be done via minor microcode changes.
Table~\ref{tab:hardware_changes} shows required hardware changes for \mbox{\Sys} and related subsections.
First, to support new enclave type, bi-enclave, SECS and EPCM entries are modified. 
Second, new instructions \textit{ESADD} and \textit{ESACCEPT} for secure channel are added, while existing instructions (\textit{EINIT}, and \textit{EEXIT}) 
are modified. 
Finally, the TLB miss handler has been changed to support different permission checks of \mbox{\Sys} with SGX.
Note that the access validation is done only when TLB miss occurs.
\Sys does not change any other components in processors and cache hierarchy.


\subsection{Memory Protection for Stockade}
\label{subsec: memory_protection}

\begin{figure}[t]
\centering
\includegraphics[width=8.5cm]{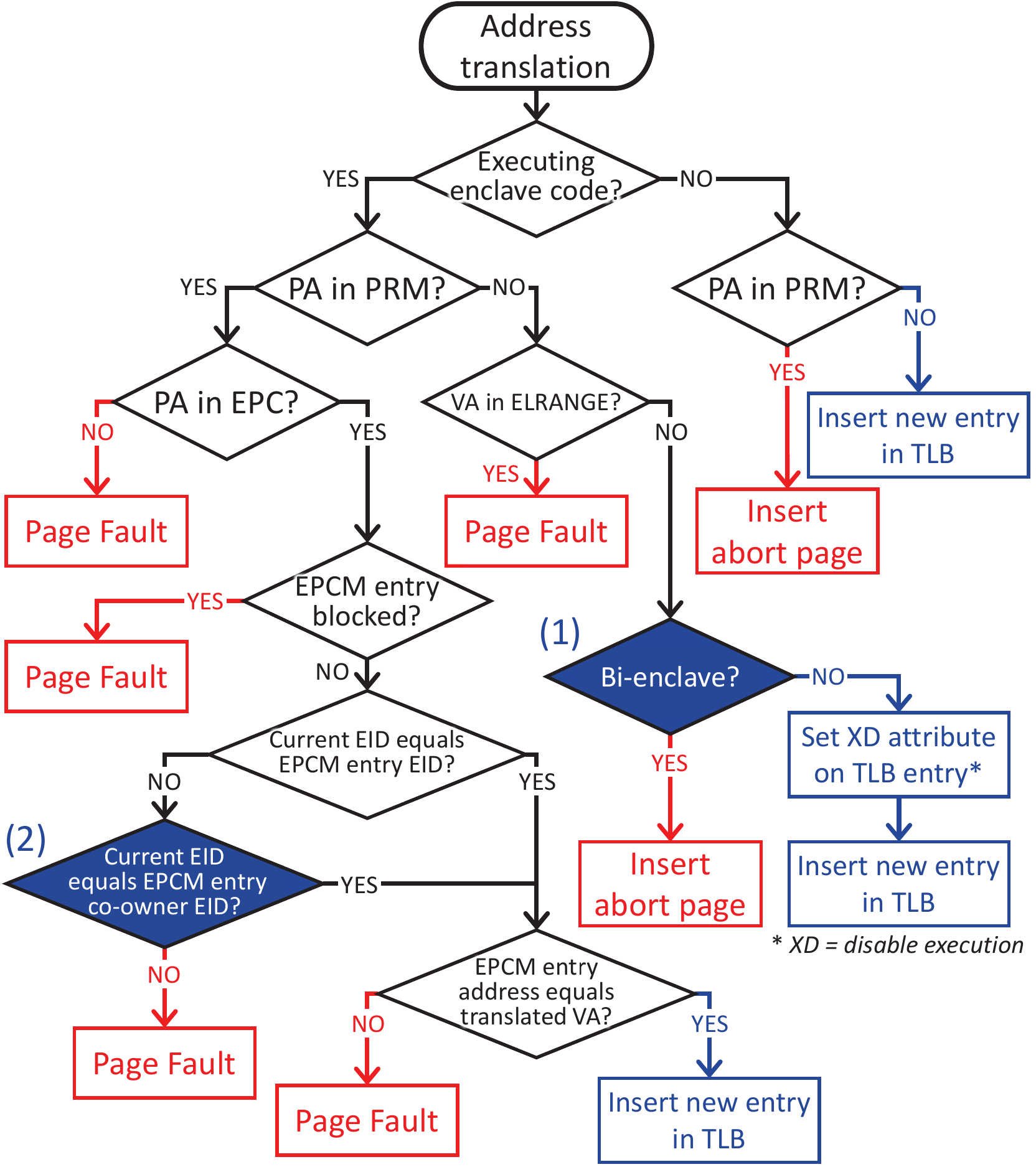}
\vspace{-0.10in}
\caption{Access control flow for \sys. Modification are painted blue on the original SGX's flow~\cite{sgx-explained}}
\label{fig:flowchart}
\end{figure}

\noindent
{\bf Access Validation:}
\Sys leverages SGX memory protection features to enable bi-directional memory protection.
Based on the SGX original memory isolation, \sys provides additional memory protection in the opposite direction.
\Sys protects the non-enclave memory context by preventing memory translation from a \se.
Figure~\ref{fig:flowchart} is the hardware flowchart for \sys's address translation. {\bf (1)} in the figure
indicates the additional memory protection added for \sys.
When an enclave is not in \se mode, memory access to a non-ELRANGE virtual address is allowed.
Thus, \sys inserts a new entry to the TLB in the same way as the original SGX.
However, when the enclave is in \se mode, the sandboxed code must not be allowed to translate to the outside memory.
\Sys inserts an abort page to cause a failure in resolving the non-ELRANGE virtual address to a physical address.
As shown in the figure, the extra access control for \se does not require any significant hardware changes;
\sys needs a minor extra condition check while handling a TLB miss.

\noindent
{\bf Control transition:}
By calling \textit{ocall}, an enclave performs a control transfer from the enclave to the non-enclave context.
To perform an ocall, a normal enclave saves its state in the protected memory, cleanses all internal CPU states
to prevent security leaks, and switches its mode into the non-enclave mode with \textit{EEXIT}.
Unlike normal enclaves, \Sys isolates a \se by disabling {\it EEXIT} instruction.
The hardware modification for {\it EEXIT} is minor since CPU only checks whether the current context is in \se mode or not
by reading flag in the SECS structure.
However, Asynchronous Enclave Exit (AEX) is still allowed even for the \se
because operating systems must handle exceptions such as page faults or interrupts.
When AEX occurs, all execution contexts are securely saved in the enclave memory, 
and \sys switches its execution mode to handle the exit events.
The event is handled by designated hardware exception handlers in processors. 
During AEX, it erases any context (secrets) that may exist in the execution state~\cite{sgx-explained}.
Therefore, the software in a bi-enclave cannot exploit AEX for escaping the sandbox.
The saved context will be restored during the next {\it EENTER} or {\it ERESUME}.
\Sys does not modify the flow of AEX from the original SGX.


\noindent
{\bf Advantages:}
Compared to the prior SW and HW+OS confinement approaches~\cite{ryoan,Chancel,AccTEE,SGXJail}, \Sys can provide hardware-supported strong isolation efficiently.
\Sys does not require any extra SW layers or compiler-based validations for the confinement, unlike the prior SW approaches.
Furthermore, \sys is more robust against Spectre-like attacks which attempt to bypass protection boundary checks of the SW approaches
because an unauthorized speculative access will incur a TLB miss and the address translation fails.
Compared to the HW+OS approaches which keep domain IDs in vulnerable page tables,
\sys keeps the critical meta-data in the secure memory region (PRM).
Therefore, the meta-data is protected from OS and from DRAM attacks including rowhammer
because any bitflips in PRM are detected by the integrity validation
of the hardware engine.




\subsection{Pairwise Secure Shared Memory}
\label{subsec:shared_epc}

For efficient communication between enclaves, \Sys introduces a new secure channel using the protected shared memory. 
The communication channel is a small piece of SGX memory exclusively shared with two enclave parties.
To share an EPC page between two enclaves, the enclave pair has to agree on the memory sharing.
Once the channel is established, the software-based encryption is not necessary for inter-enclave communication.
In \sys, establishing the shared channel is done by hardware. 
Moreover, the channel can avoid even the encryption by hardware, if the contents fit in the CPU caches for efficient data exchange.



EPCM in SGX contains the EPC mapping information to validate translation.
To support the shared memory, the EPCM entry should be modified to include the sharer information.
In addition to the owner enclave's SECS address, an shared EPCM entry includes a single co-owner enclave's SECS.
The SECS address is represented physical page number related to start address of EPC~\cite{sgx-explained}.
\Sys extends EPCM in the same way for the co-owner.
In \sys, a single EPC page can be shared only between two enclaves.
Figure~\ref{fig:flowchart} shows the hardware extension for the ownership checking during a TLB miss at EPC.
In {\bf (2)} of the figure, when a memory access occurs to EPC,
\sys checks the corresponding EPCM entry and verifies the owner enclave.
The EPCM entry has at most two enclaves as its owner and co-owner, and
thus only two different enclave contexts can access the EPC page.

\noindent
{\bf Sharing EPC memory:}
\Sys provides a new user level instruction, {\it ESADD}, to share a EPC page with a co-owner enclave.
The instruction takes the EPC page address of owner enclave and the ID of co-owner enclave as inputs.
When {\it ESADD} is invoked, the SGX hardware zeros the page and blocks any access to the page until the corresponding co-owner invokes {\it ESACCEPT}.
{\it ESACCEPT} performs TLB synchronization to remove old mappings in TLB,
and write co-owner down on corresponding EPCM.
This step is similar to the dynamic EPC expansion instructions in SGX2~\cite{sgx2}.

Once a shared memory is established, two enclaves initiate the local attestation step.
Each enclave's digest ({\it MRENCLAVE}) is passed through the newly established shared memory.
The digest uniquely identifies each enclave because it records all the enclave contents
(page contents, related position, security flags)~\cite{sgx-explained}.
If both enclaves verify each other successfully, then they finalize the channel establishment.
Otherwise, \sys destroys the channel.

\noindent
{\bf Communication via APIs:}
The communication API is similar to ecall and ocall, but all the arguments are secured.
When an enclave module invokes the API, the module performs sanitizing and marshalling
all the parameters to the structure allocated in shared EPC.
After that, \sys copies the parameters into callee's private memory to prevent possible TOCTOU(time-of-check-time-of-use)
attacks~\cite{TOCTOU} and performs de-marshalling to execute a callee's function.
Because caller and callee belong to different enclaves, CPU state flush is not necessary.

\subsection{Sandbox Monitor}
\label{subsec:monitor}
A monitor enclave executes a software reference monitor which verifies system calls and returns values. 
For legitimate system calls, the monitor enclave executes those on behalf of \mbox{\ses}.
The monitor can execute system calls to the kernel or can leverage Intel-provided C standard libraries~\cite{developer_guide}.

\noindent
\textbf{Policy Loading: }
The monitor enclave reads a policy definition file for the \se
which is mutually agreed and shared in advance by the application module provider and
the cloud provider.
To verify the policy file is correctly loaded, both a \se and an OS can query the monitor enclave
to obtain the digest of the file.
\Se can request the query with the key made during local attestation.
\Se checks the digest value matches with its own policy digest
to make sure the file is not manipulated by the OS.
The monitor enclave opens an upcall interface only for serving the digest query from the OS.
This is similar approach to Intel's attestation to verify an enclave's identity.

\begin{lstlisting}[caption={Example policy file}, captionpos=b, label={lst:policy}]
SYS_NUM ACTION
0        0      // read     ALLOW
1        2      // write    NOTIFY
2        1      // open     LOG
42       5      // connect  KILL
43       3      // accept   TRAP

BLACKLIST   0  "/path/to/top/secret*"
WHITELIST   2  "/path/to/no/secret/[a-z_\-\s0-9\.]"
BLACKLIST  43  "112.233.0.0/16"
\end{lstlisting}

\noindent
\textbf{Monitor as mediator: }
Listing ~\ref{lst:policy} shows an example policy definition file supported in our system.
Similar to seccomp-bpf~\cite{seccomp}, the monitor enclave filters each system call with system call ID
and its arguments based on fine-grained privileges to system resources through a blacklist and a whitelist.
To speed up the syscall filtering, the sandbox monitor can adopt an action-based policy.
If an action specifies KILL, the monitor enclave sends a request to the kernel to terminate the \se.
On NOTIFY, the monitor makes a notification to the kernel and continue the execution.
When the action is LOG, it writes encrypted logs of the system call to a file.
On TRAP, the monitor runs a customized logic (e.g. sends a message to the module provider).

When a system call returns, the monitor enclave verifies the return value from the kernel
to prevent Iago attacks.
For example, 
Since most system calls return boolean or integer type~\cite{PANOPLY}, the monitor can check
whether the return values belong to a proper range of values.
\sys checks whether \textit{futex}, \textit{locks}, and \textit{semaphore} are not shared between \se and untrusted world.
For a system call that returns a descriptor or reference (e.g. {\it open}, {\it socket}),
the monitor enclave keeps it in its memory so that the returned descriptor is not substituted and reused.
In addition, \sys is resilient to pointer misuses since the reference is not accessible by a \se.


\noindent
\textbf{Trusted accounting: }
As discusses, the monitor enclave builds mutual trust.
The monitor enclave is isolated from both the user-provided \se and the host operating system,
so it works in the neutral area where the \se and kernel can trust.
This model makes new functionalities deployed in the monitoring enable other than
system call monitoring.
One use case is a trusted resource accounting system for function-as-a-service.
The cloud resource usage accounting needs to be verified by both users and provider~\cite{HRA}.
For example, the monitor enclave can record tamper-proof evidence (e.g., log file)
for network and file usages because all accesses (system calls) to the resource must pass
through the monitor enclave. Therefore, the monitor enclave can log resource requests
from each \se, and neither a \se nor the host OS cannot modify the
log contents.

\begin{table}[t]
\centering
\resizebox{8cm}{!}{
\small
\begin{tabular}{lrrrr}
\toprule
\textbf{Benchmark}						& \textbf{SGX enabled Lib} & \textbf{\begin{tabular}[c]{@{}c@{}} \# of allowed\\ interfaces\end{tabular}} & \textbf{\begin{tabular}[c]{@{}c@{}}Modified\\LOC\end{tabular}} \\
\midrule
\multirow{1}{*}{NBench}					& None  							& 0		& 8	 \\ 
\multirow{1}{*}{SSL Server} 			& OpenSSL~\cite{sgx-openssl}		& 18 	& 8  \\
\multirow{1}{*}{File I/O bench}			& Protected FS~\cite{protected_fs}	& 19	& 12 \\
\multirow{1}{*}{YCSB (SQL)} 			& SQLite~\cite{sgx-sqlite}			& 12	& 56 \\
\multirow{1}{*}{ML benchmark}			& LibSVM~\cite{libsvm}				& 7		& 8	 \\
\multirow{1}{*}{FTPS Server}			& OpenSSL \& Protected FS 			& 37	& 20 \\
\bottomrule
\end{tabular}
}
\caption{Benchmarks for evaluation.}
\label{tab:syscall_allowed}
\vspace{-0.1in}
\end{table}

\renewcommand{\arraystretch}{0.9}
\begin{table}
\centering
\resizebox{8.8cm}{!}{%
\small
\begin{tabular}{@{}lllc@{}}
\toprule
Type & Attacker & Target  & Section
\\
\midrule
Read / Write / Execute & OS & \Se, Monitor &$\S$\ref{subsec: memory_protection}
\\
Read / Write / Execute & \Se & Other \Se, Monitor &$\S$\ref{subsec: memory_protection}
\\
Read / Write / Execute  & Monitor & \Se &$\S$\ref{subsec: memory_protection}
\\
Read / Write / Execute  & \Se & Outside sandbox &$\S$\ref{subsec: memory_protection}
\\
Transfer control  & \Se & Other \Se, Monitor &$\S$\ref{subsec: memory_protection}
\\
Transfer control & \Se & Outside sandbox &$\S$\ref{subsec: memory_protection}
\\

Establish a connection & OS & \Se, Monitor  &$\S$\ref{subsec:shared_epc}
\\
Evasdrop / Modify & OS & Shared Channel &$\S$\ref{subsec:shared_epc}
\\ 
Known Iago attacks & OS & \Se, Monitor &$\S$\ref{subsec:monitor}
\\
\bottomrule
\end{tabular}%
}
\caption{Summary of security analysis of \Sys}
\label{tab:sec_analysis}
\vspace{-0.1in}
\end{table}
\renewcommand{\arraystretch}{1.0}

%

\section{Discussion}
\label{sec:discussion}

\noindent
\subsection{Development in \sys}
\vspace{-0.05in}
Like other SGX-based sandboxing~\cite{ryoan}, applications using \sys are
compartmentalized based on protection domains. \sys executes each protection domain
in separate \ses.
Module providers should specify a policy
for their modules as discussed in Section~\mbox{\ref{subsec:monitor}}.
To communicate among \ses, \sys provides APIs in the SGX SDK.
Application developers simply replace the existing communication APIs (e.g. ecall/ocall) with \sys's APIs
so that the module providers do not need to care about new interfaces.
Table~\ref{tab:syscall_allowed} lists benchmark with allowed interfaces and modified LOC.
When the application is already ported into SGX, porting to \sys only requires few lines for initial setup.
Most of porting would be done in Makefile, which would be provided by cloud provider.
If the application involves mutually untrusted modules written by multiple parties,
developer has to map each untrusted module to a \se.

A case requiring developers' porting efforts is when an application's SGX module
is written to communicate with non-enclave code (e.g., ocall to untrusted libraries in non-enclave mode).
Because \mbox{\sys} does not allow such communications, developers must port the non-enclave code to run inside another bi-enclave.
In addition, each bi-enclave must not access over its sandbox limit to avoid abort page. 

\subsection{Security Analysis}
\vspace{-0.05in}
\autoref{tab:sec_analysis} summarizes the security analysis of \sys.
An attacker tries to break isolations of \se by compromising or launching a \se.
However, the attacker cannot run or access untrusted context even with \textit{ret}, \textit{jmp} and \textit{EEXIT}
as described in~$\S$\ref{subsec: memory_protection}.
The compromised \se cannot transfer its control to the monitor enclave or another \se because they are separate enclaves.
Also, the compromised \se cannot create a shared memory with an arbitrary \se
because every shared memory establishment is verified with the attestation.
On the other hands, malicious system software cannot eavesdrop or hijack
communication to mount the man-in-the-middle attacks~\cite{PANOPLY}.
\Sys allows the enclave-to-enclave communication channel only via the SGX-protected
memory, so the OS or hypervisor is not able to access the communication channel
unlike what the original SGX does.
The possible attack surfaces are syscall 
interfaces for the monitor enclave. We implemented
known Iago attack protections by checking the file descriptor from syscall~\cite{Overshadow,Inktag} and POSIX semaphore invocations~\cite{Sego}.

\noindent
\textbf{Limitation:}
An attacker may subvert the entire application by gathering code-reuse gadgets one by one across multiple modules
and exploiting vulnerable APIs between them.
As we mentioned in section~\ref{subsec:threatmodel} we leave this our limitation.
\section{Evaluation}
\vspace{-0.05in}
\label{sec:evaluation}
\begin{table}[t]
\centering
\resizebox{8cm}{!}{
\small
\begin{tabular}{lrrr}
\toprule
& Hardware mode & Simulation mode \\\midrule
SGX-NBench(geomean)~\cite{sgx-nbench} & 6.0 & 6.4 \\
SGX ecall / ocall (switchless) & 2283.8 / 3748.5 & 3110.3 / 3783.7 \\
\Sys inter-enclave call & - & 4930.5 \\\bottomrule
\end{tabular}
}
\caption{Hardware and simulation mode performace comparison (1000 iterations/sec)}
\vspace{-0.1in}
\label{tab:emulation_justification}
\end{table}


\begin{figure}[t]
\centering
\includegraphics[width=8.5cm]{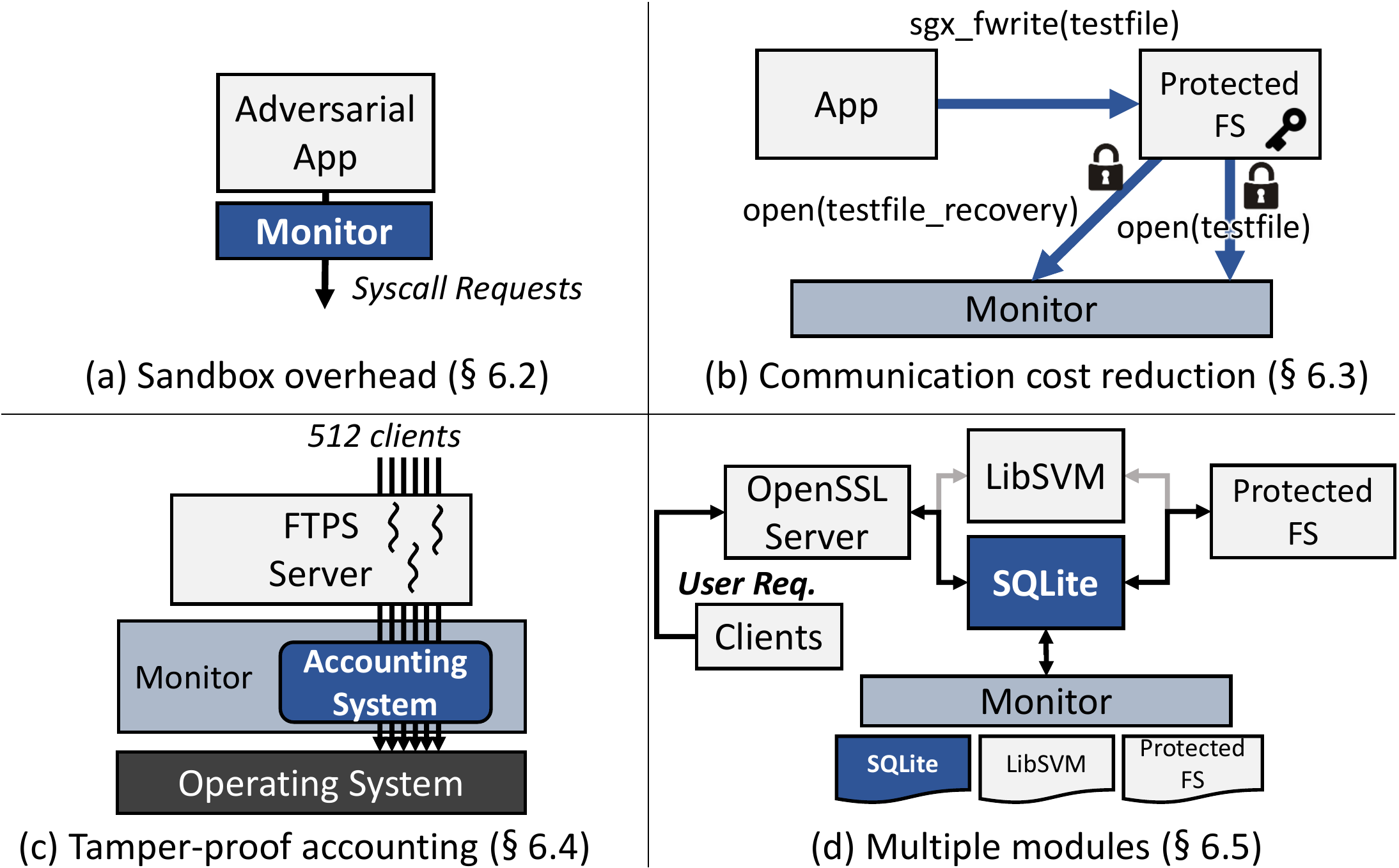}
\vspace{-0.20in}
\caption{Evaluation scenarios}
\label{fig:casestudy}
\vspace{-0.05in}
\end{figure}

\begin{figure*}[t]
\centering
\includegraphics[width=\textwidth]{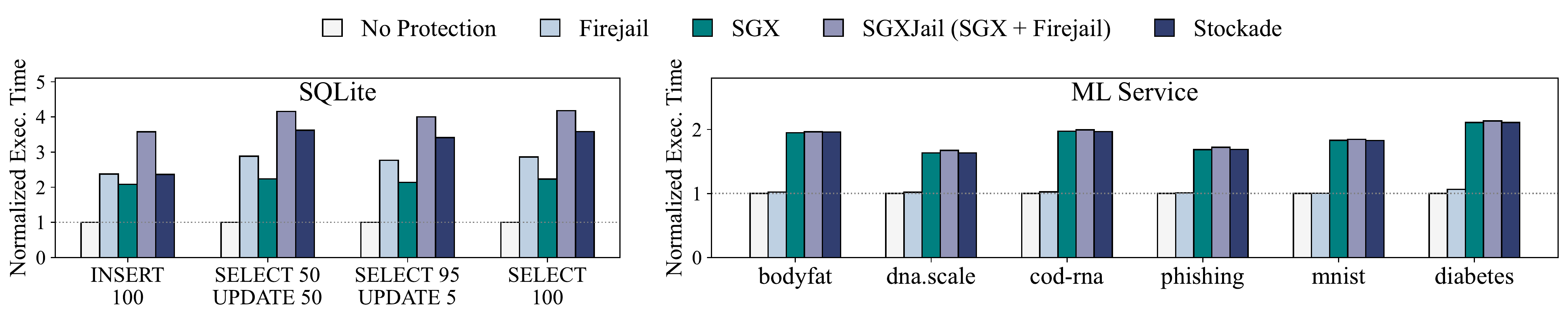} \\
\vspace{-0.1in}
\caption{Comparison of execution times between applications run on several secure systems including \sys}
\label{fig:perf}
\end{figure*}



\subsection{Methodology}

\noindent
\textbf{Environment:}
We evaluate \sys in servers consisting of Intel CPU i7-7700, 64GB DDR4 DRAM, and Ubuntu 16.04 with Linux kernel 4.13.0.
To add new hardware features, we use the simulation mode in Intel SGX driver and SDK version 2.2.
The simulation mode supports SGX APIs, trusted libraries, and emulation for SGX instructions~\cite{sgx-simulation}.
Table~\ref{tab:emulation_justification} shows performance comparisons between the hardware mode and the simulation mode.
To capture the effect of TLB shootdown, \sys sends ioctl to SGX driver.
The driver runs \textit{mov cr3, cr3}, which flushes TLB of the process.

\noindent
\textbf{\sys features:}
\sys hardware features are implemented mostly in SDK and Driver.
Disabling ocall from \se is done by modifying emulated {\it EEXIT} instruction
as described in $\S$\ref{subsec: memory_protection}.
We modified \textit{Edger8r} in SDK to generates APIs from Enclave Defined Language (EDL) format.
Based on the format, the \textit{Edger8r} per-API data structure for type and boundary checking.
For example, pointer arrays require the number of elements to be passed for marshaling.
The generated APIs are linked to enclave modules at compile time.

\subsection{Sandbox Overhead}
\label{subsec:sandbox_overhead}
\vspace{-0.05in}

In this section, we measure sandboxing overhead of \sys
compared to software-based approaches that use process isolation and binary instrumentation.
\autoref{fig:casestudy} (a) shows the evaluation scenario.

\noindent
\textbf{Comparison to process-based filtering:}
SGXJail is a sandbox that isolates an enclave instance in a separated process
confined by seccomp filters~\cite{seccomp}. Because SGXJail is not open-sourced,
we emualte SGXJail in our platform.
To reproduce SGXJail's performance,
we use Firejail~\cite{firejail} which leverages Linux namespace and seccomp
to provide system call interposition.
We compared \sys with four control groups:
no protections from SGX nor sandbox ({\tt No Protection}),
software-based sandbox ({\tt Firejail}),
SGX enclave ({\tt SGX}),
and SGX enclave with the software sandbox ({\tt SGXJail}).

Figure~\ref{fig:perf} shows normalized execution time running SQLite and ML Service.
We evaluate SQLite as an I/O-intensive benchmark. SQLite runs a set of queries generated by YCSB~\cite{YCSB}.
Each set contains 10,000 queries of INSERT, SELECT, and UPDATE according to the uniform key distribution in different ratios.
A higher ratio in SELECT queries degrades the performance over INSERT/UPDATE
since SELECT generates more syscalls than others to traverse a database.
{\tt SGXJail} shows the slowest performance: 1.83$\times$ slower on average compared to {\tt SGX}
due to frequent system call monitoring by {\tt Firejail}.
Meanwhile, {\tt \Sys} is only 1.49$\times$ slower on average than {\tt SGX}
as {\tt \Sys} passes syscall requests without costly IPC.
As a result, {\tt \Sys} shows 18.6\% better performance over {\tt SGXJail}
while providing stronger hardware-based isolation.
In ML services, {\tt SGX}, {\tt SGXJail}, and {\tt \Sys} are similar in speed, less than 3\%,
as ML inference is CPU-intensive and seldom invokes syscalls.

\begin{figure}[t]
\centering
\includegraphics[width=8.5cm]{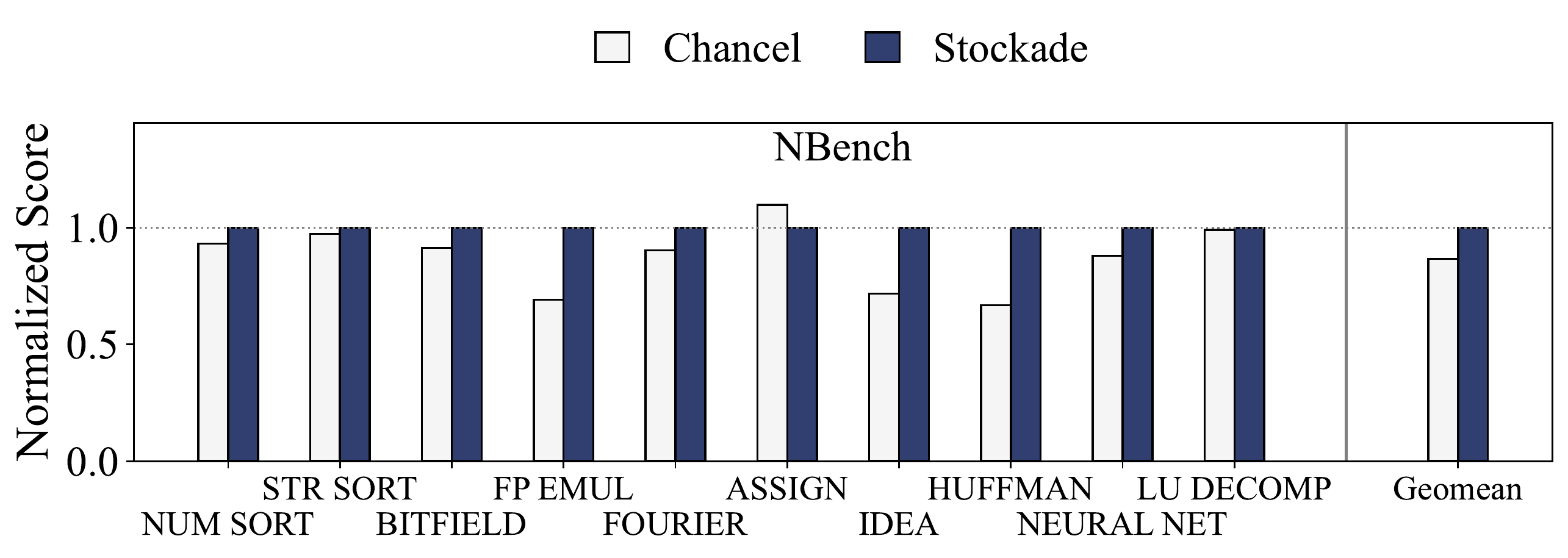}
\vspace{-0.20in}
    \caption{Normalized performance of NBench.}
\label{fig:vschancel}
\end{figure}


\noindent
\textbf{Comparison to binary instrumentation:}
We compare \sys with Chancel~\cite{Chancel}, a software-based binary instrumentation approach
for bi-directional isolation like \sys.
We run {\tt NBench} which consists of ten benchmarks exposing CPU, FPU, and memory capabilities~\cite{nbench}.
For fair comparision, we run \sys with clang-4.0 with -O0 option as the same configure of Chancel.
Figure~\ref{fig:vschancel} shows normalized performance degradations.
We normalize performance by non-confined baseline.
In {\tt NBench}, Chancel shows 12.3\% performance degradations on average over its baseline.
The overhead is from Chancel's binary instrumentation which adds additional instructions (+23.5\%).
However, \sys runs {\tt NBench} on hardware confined areas and doesn't degrade performance
compared to the baseline because {\tt NBench} doesn't communicate to other modules,
thereby it does not incur IPC and system call monitoring overhead in the monitor enclave.

\subsection{Communication cost reduction with \Sys}
\vspace{-0.1in}

\label{eval:comm}
\begin{figure}[t]
\centering
\includegraphics[width=8.5cm]{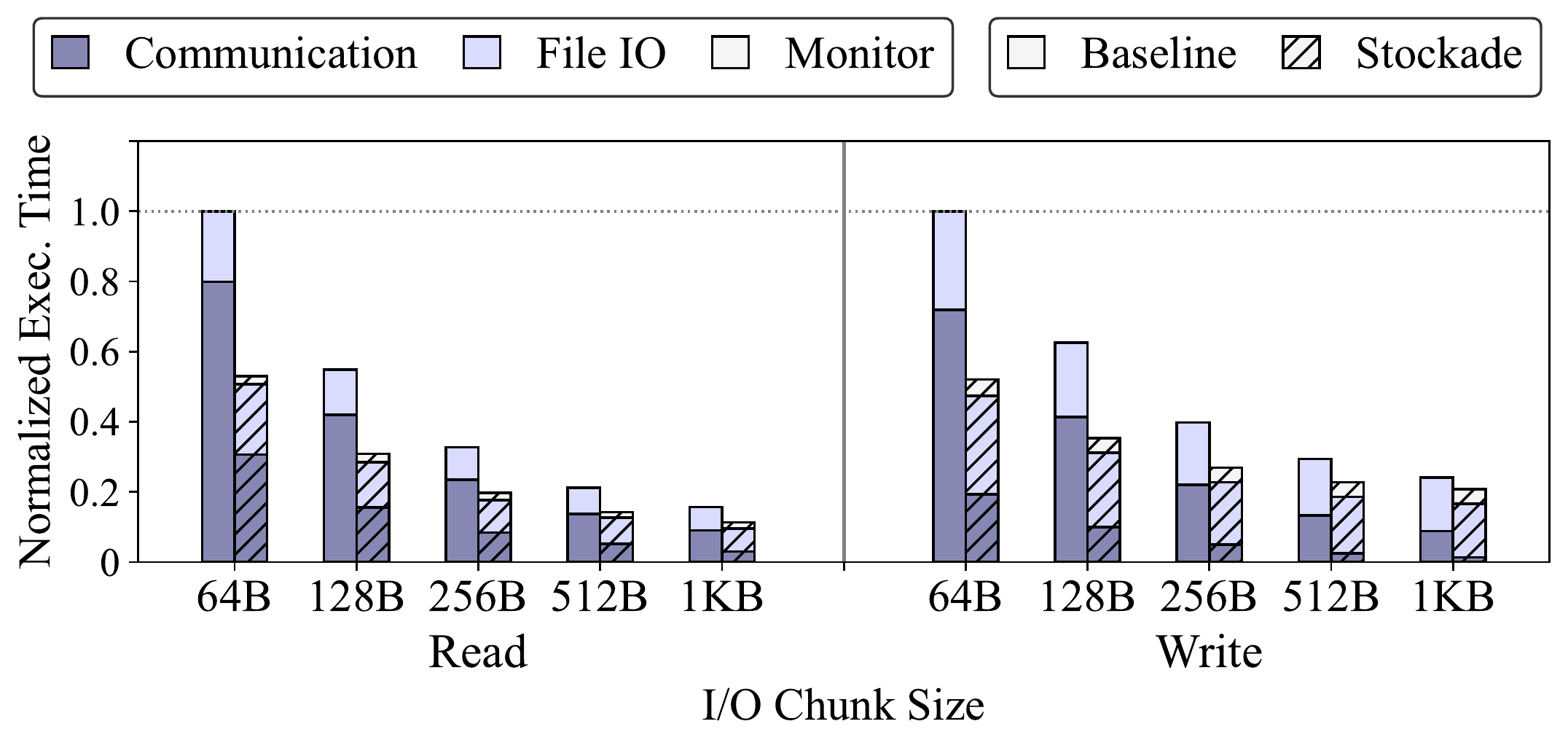}
\vspace{-0.25in}
    \caption{Execution time breakdown in file I/O scenario}
\label{fig:fileio}
\end{figure}

\noindent
In this section, we compare the communication cost of \sys (HW) to
SW-based approach.
\autoref{fig:casestudy} (b) represents the scenario.
We develop Protected FS which provides integrity and confidentiality protection of files.
A module runs the Protected FS, and a communicating module uses the Protected FS to secure its file I/O.
The communication is done via the \sys's secure channel, so the content of the file and messages are secured.
For the same guarantee, We implement {\tt Baseline} which uses software encryption for secure communication,
but it does not confine modules.

\autoref{fig:fileio} presents normalized execution time in various chunk sizes.
The execution time is normalized to when the chunk size is 64B of {\tt Baseline}.
To observe communication cost clearly, we perform file I/O on mounted tmpfs (DRAM backend).
We breakdown its performance by three factors.
{\tt File I/O} indicates the time taken for file APIs, and
{\tt Communication} shows execution time for all communications between modules including message serialization and encryption.
{\tt Monitor} is the overhead of \sys monitor.
\sys's HW-based communication effectively
saves the communication cost than SW-based approaches.
As consequence, {\tt \Sys} shows up to 1.38$\times$ faster in {\tt read}, 1.16$\times$ faster in write comparing to {\tt Baseline}
when the chunk size is 1KB.
This speedup is caused by the elimination of costly software encryption and decryption.
When the chunk size is 64B, {\tt \Sys} shows up to 1.89$\times$ faster in read, 1.92$\times$ faster in write comparing to {\tt Baseline}.
Fine-grained file I/O operations causes more frequent communication overhead between modules,
thus it stresses the communication cost.
\subsection{Tamper-proof Accounting System}
\vspace{-0.10in}

\begin{figure}[t]
\centering
\includegraphics[width=8.5cm]{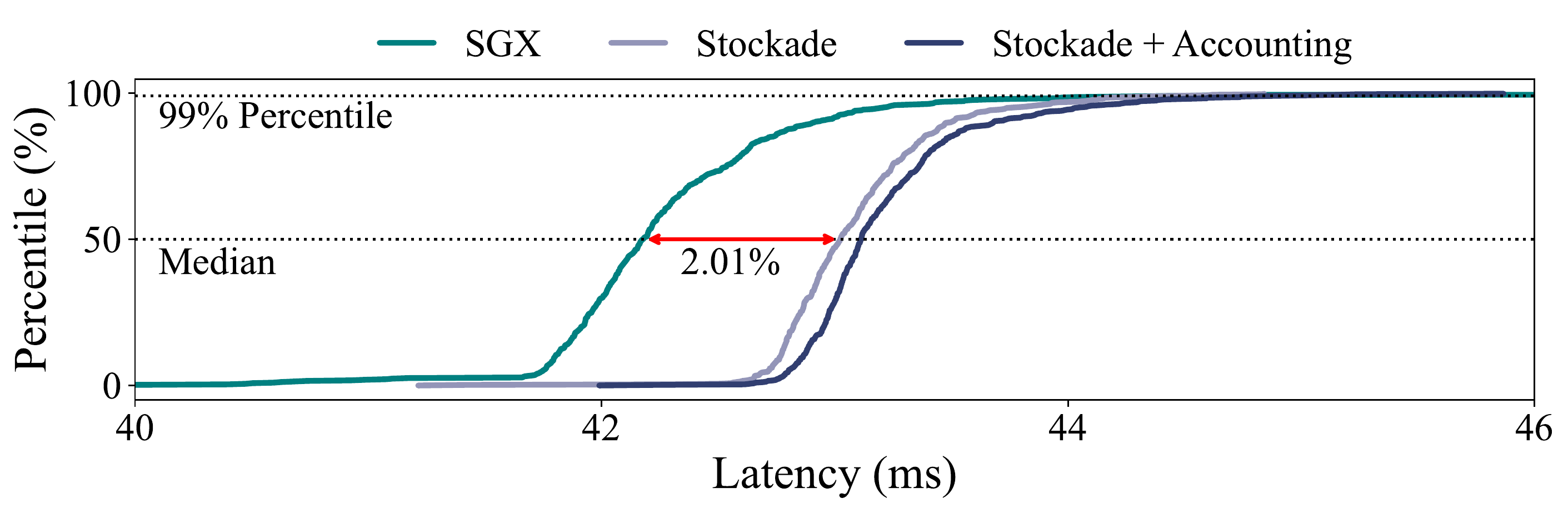}
\vspace{-0.20in}
    \caption{Latency distribution of FTPS requests}
\vspace{-0.05in}
\label{fig:res-account}
\end{figure}

In a cloud system, multiple tenants compete to use the resource from limited hardware.
Trusted tamper-proof accounting systems provide useful feature to securely manage resources
across the tenants \cite{HRA, T-lease, AccTEE}.
\Sys can provide such a system using a monitor enclave as shown in figure~\autoref{fig:casestudy} (c).
Because the monitor enclave intervenes all the system calls from \ses,
it can account every stat of system calls (e.g. file I/O access, network request, memory consumption).
\sys provides a trustworthy accounting reports
that the service provider can verify through attestation.

To demonstrate the scenario, we implement the following accounting system in the monitor enclave.
We spawn 512 clients; each of them sends a request for a 1MB file to secure FTP server.
The server takes the request and sends back corresponding files to clients.
Handling the requests, the monitor enclave logs per-request resource
consumption in its secure memory.
\autoref{fig:res-account} shows the latency distribution
of the requests with three different systems: {\tt SGX}, {\tt \Sys}, {\tt \Sys+ Accounting}.
The median latency of {\tt \Sys} is only 2\% slower than {\tt SGX} and
the {\tt \Sys+ Accounting} shows negligible overhead in median and tail latencies.
This implies that mutually trustful secure accounting can be achieved without large overhead via \sys.

\subsection{Secure Services with Multiple Modules}
\vspace{-0.1in}

To evaluate combined benefits of \sys, we build a secure query server containing DB and ML services.
\autoref{fig:casestudy} (d) describes the system; each service consists of multiple modules and they are isolated to each \se.
The modules attests to each other and establishes secure communication channels at first.
The monitor enclave reads a system call policy that specifies the least privilege of each module.
For example, SQLite is not allowed to use network-related system calls such as {\it connect} or {\it send}.
Each service uses confined third-party modules for secure network communication {\tt (SSL Server)}
and safe file management {\tt (Protected FS)}.
Whenever a client sends security-sensitive data to DB service via {\tt SSL Server},
SQLite module asks the protected file system to store or load the data.
{\tt Protected FS} has its own encryption key which is not accessible from {\tt SSL Server} or SQLite module.
In addition, the LibSVM module handles a prediction with the inference model
stored in the local file system encrypted by {\tt Protected FS}.
Even though {\tt SSL Server} is compromised,
the inference model cannot be stolen due to \sys's isolation.

\begin{figure}[t]
\centering
\includegraphics[width=8.5cm]{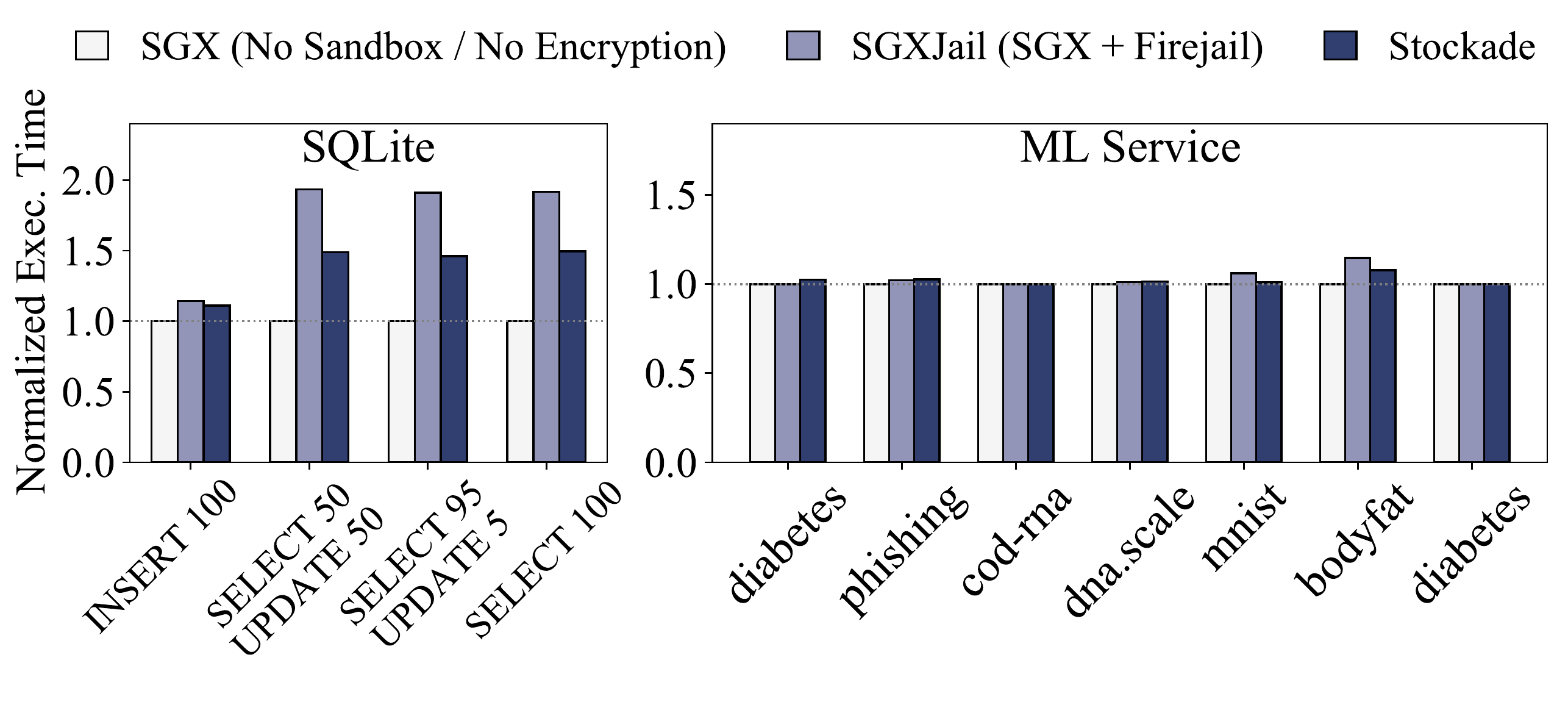}
\vspace{-0.30in}
    \caption{Normalized execution time of distributed query server scenarios}
\label{fig:securequeryserver}
\end{figure}

\autoref{fig:securequeryserver} shows the performance of the two services.
{\tt SGX} indicates each module runs in an enclave and communicates
with each other via unprotected channel without encryption. 
Each enclave is not confined, so it can perform any system calls.
However, in {\tt SGXJail} and {\tt \Sys}, every system call has verified. 
The sandbox overhead of {\tt \Sys} incurs a slowdown as shown in $\S$\ref{subsec:sandbox_overhead},
but the efficient hardware-based encryption amortizes the performance degradation.
As a result, for I/O-intensive SQLite,
{\tt \Sys} shows 38.9\% overhead compared to {\tt SGX} and 19.5\% better to {\tt SGXJail} on average.
Despite communication overhead from adding monitor enclave, \mbox{\Sys} outperformed {\tt SGXJail} by leveraging hardware encryption.
For ML service, the performance among the three models are similar, but some overhead is added in {\tt SGXJail} and {\tt \sys} due to secured communication.



\section{Conclusion}
\label{sec:conclusion}

This paper explores a new extension model, \sys, for SGX to support distributed sandboxing.
With a minor change in SGX, \sys provides strong sandboxing. In addition,
it allows the mutually trusted monitor enclave between the user \se and the operating system,
by filtering system call requests and validating return values.
The performance results show the viability of \sys.
In multi-module, \mbox{\Sys} shows an average 19.5\% speedup for SQLite, and 1.4\% speedup for ML service over the SW sandbox approach.

\section{Acknowledgements}
This work was supported by Institute
for Information \& communications Technology Promotion
(IITP2017-0-00466). The grant is funded by the Ministry of
Science and ICT, Korea.
This work was also partly supported
by Samsung Electronics Co., Ltd. (IO201209-07864-01).

\bibliographystyle{plain}
\bibliography{references}

\end{document}